\documentclass{aa}
\usepackage{graphics}

\begin{document}
\title{Thermal infrared observations of near-Earth asteroid
       \object{2002~NY40}
       \thanks{Based on observations collected at the European
               Southern Observatory, Chile;
               ESO, No.\ 69.C-0152}}
\author{T.\ G.\ M\"uller\inst{1}
	\and
	M.\ F.\ Sterzik\inst{2}
        \and
	O.\ Sch\"utz\inst{3}
	\and
	P.\ Pravec\inst{4}
	\and
	R.\ Siebenmorgen\inst{5}}

\authorrunning{M\"uller et al.}
\titlerunning{2002~NY40 in the thermal infrared}

\offprints{T.\ G.\ M\"uller}

\institute{%
    Max-Planck-Institut f\"ur extraterrestrische Physik,
    Giessenbachstra{\ss}e, 85748 Garching, Germany;
    \email{tmueller@mpe.mpg.de}
 \and
    European Southern Observatory, Casilla 19001, Santiago 19, Chile;
    \email{msterzik@eso.org}
 \and
     Max-Planck-Institut f\"ur Astronomie, K\"onigstuhl 17,
     69117 Heidelberg, Germany;
     \email{schuetz@mpia-hd.mpg.de}
 \and
    Astronomical Institute, Academy of Sciences of the Czech Republic,
    Ondrejov, CZ-25165, Czech Republic;
    \email{ppravec@asu.cas.cz}
 \and
    European Southern Observatory, Karl-Schwarzschildstr.\ 2,
    85748 Garching, Germany;
    \email{rsiebenm@eso.org}}

\date{Received / Accepted}

\abstract{We obtained N-band observations of the Apollo asteroid \object{2002~NY40}
          during its close Earth fly-by in August 2002 with TIMMI2 at the ESO 3.6\,m
          telescope. The photometric measurement allowed us to derive a radiometric
          diameter of 0.28$\pm$0.03\,km and an albedo of 0.34$\pm$0.06 through the
	  near-Earth asteroid thermal model (NEATM) and a thermophysical model (TPM).
	  The values are in agreement with results from radar data, visual and
          near-IR observations. 
	  In this first comparison between these two model approaches
	  we found that the empirical NEATM beaming parameter
	  $\eta=1.0$ corresponds to a thermal inertia values of about
	  100\,$\mathrm{J\,m^{-2}\,s^{-0.5}\,K^{-1}}$ for a typical range of
	  surface roughness, assuming an equator-on viewing angle. 
          Our TPM analysis indicated that the surface of \object{2002~NY40} consists
	  of rocky material with a thin or no dust regolith.
	  The asteroid very likely has a prograde sense of rotation with
          a cold terminator at the time of our observations.
          Although both model approaches can fit the
	  thermal spectra taken at phase angles of 22$^{\circ}$ and 59$^{\circ}$,
	  we did not find a consistent model solution that describes all pieces of 
	  photometric and spectroscopic data.
	  In addition to the \object{2002~NY40} analysis, we discuss the possibilities
          to distinguish between different models with only very few photometric and/or
	  spectroscopic measurements spread over a range of phase angles.
  \keywords{Minor planets, asteroids -- Radiation mechanisms: Thermal --
            Infrared: Solar system}
}

\maketitle
%_______________________________________________________________________

%%%%%%%%%%%%%%%%%%%%%%%%%%%%
%     Introduction         %
%%%%%%%%%%%%%%%%%%%%%%%%%%%%
\section{Introduction}

  The Apollo asteroid \object{2002~NY40} was discovered by
  LINEAR\footnote{Lincoln Near Earth Asteroid Research project}
%  is a MIT Lincoln Laboratory program funded by the United States
%  Air Force and NASA.} on July 14th, 2002
  (MPEC 2002-O17\footnote{http://cfa-www.harvard.edu/mpec/K02/K02O17.html}).
  It passed the Earth at 0.0035\,AU, about 1.3 times the distance
  to the Moon, on August 18th, 2002. This event was closely
  followed by many amateur and professional observers at a large
  range of wavelengths from visible to radar. 

  Pravec et al. (\cite{pravec04}) found from 21 different observing
  sessions a main lightcurve period
  of $P_1=19.98\pm0.01$\,h with indications for a non-principal axis
  rotation with a non-unique but likely second period of $P_2=18.43\pm0.01$\,h.
  The lightcurve shows sharp bends, indicating the presence of large
  non-convex shape features. Its mean absolute magnitude 
  is $H=19.23\pm0.2$\,mag, with variations between 18.9\,mag
  for the lightcurve maximum and 19.8\,mag at the secondary minimum
  (see Bowell et al.\ (\cite{bowell89}) for a definition of the H-G system).
  
  Howell et al.\ ({\cite{howell03}) presented radar images which show
  that \object{2002~NY40} looks like two spheroidal units joined together,
  but high resolution optical imaging showed no obvious evidence for
  a binary structure
  (H. Mathis\footnote{http://www.noao.edu/outreach/press/pr02/pr0207.html}).
  
  0.3 to 4.0\,$\mu$m spectroscopy revealed absorption bands at 1 and 2\,$\mu$m
  due to olivine and pyroxene (Rivkin et al.\ \cite{rivkin03}).
  The striking similarity with LL6 chondrites, a subset of ordinary
  chondrites, indicates that \object{2002~NY40} could be a progenitor
  of LL6 meteorites. Based on the visible spectrum, Rivkin et al.\ (\cite{rivkin03})
  classified \object{2002~NY40} as Q-class asteroid, which is connected to a
  moderately high albedo of 20 - 30\,\% (Clark et al.\ \cite{clark02}).

  First size estimates were in the order of 700\,m to 800\,m
  diameter$^{3,}$\footnote{http://science.nasa.gov/headlines/y2002/30jul\_ny40.htm},
  based on magnitude estimates from early astrometric
  measurements.
  Later on, adaptive optics systems lowered this number to an
  upper limit of 400\,m at the time of the
  observation\footnote{http://www.ing.iac.es/PR/press/ing32002.html}.
  Radar data lead to a size estimate of about 250$\times$420\,m
  (E.\ Howell, personal communication) consistent with radiometric results
  (Rivkin et al.\ \cite{rivkin03}) from 2.5\,$\mu$m observations.
  They also gave a possible albedo range of 0.15-0.25 with a slight
  preference for the higher value.
  
  The radiometric method of diameter and albedo determination from
  thermal radiation measurements goes back to the early 1970s
  (references are given in e.g., Morrison \& Lebofsky \cite{morrison79}).
  A widespread version of this technique, the Standard Thermal
  Model (STM), was published by Lebofsky et al.\ (\cite{lebofsky86})
  and discussed by e.g., Lebofsky \& Spencer (\cite{lebofsky89}).
  The STM uses an empirical beaming parameter $\eta$, which
  was introduced to adjust the sub-solar surface temperature and
  to account for non-isotropic heat radiation (Lebofsky et al.\ \cite{lebofsky89}).
  The STM is strongly connected to the IRAS minor planet catalogue
  (Tedesco et al.\ \cite{tedesco02}), with more than 2000 diameter
  and albedo values derived from thermal observations.
  Morrison (\cite{morrison76}) and Cruikshank \& Jones (\cite{cruikshank77})
  applied this technique for the first time to near-Earth asteroids (NEA).
  However, very early on Lebofsky et al.\ (\cite{lebofsky79}) and
  Veeder et al.\ (\cite{veeder89}) encountered
  problems when applying the STM to NEAs. They found in several
  cases albedos which were too high to be consistent with previous
  taxonomic classifications.

  In recent years, Harris (\cite{harris98}) and Delb\'o \& Harris
  (\cite{delbo02}) focused on the improvement of the radiometric technique
  for NEAs, resulting in the NEA thermal model (NEATM).
  The NEATM also incorporates a beaming parameter $\eta$, but now
  it is used as a variable which has to be adjusted to produce a
  fit to the spectral data. 
  Harris (\cite{harris98}), Harris et al.\ (\cite{harris98a}) and
  Delb\'o et al.\ (\cite{delbo03}) applied the NEATM successfully to
  thermal N- and Q-band observations of a large number of NEAs.
  Harris \& Davies (\cite{harris99}) presented spectrophotometric
  observations of 3 NEAs and discussed the implications for their
  physical characterisation.

  A thermophysical model (TPM) was developed by Lagerros
  (\cite{lagerros96}; \cite{lagerros97}; \cite{lagerros98}).
  A number of physical processes were introduced in this model.
  Instead of the empirical correction parameters, like the phase
  angle correction in the STM or the beaming parameter
  in the STM and NEATM, the TPM takes the true illumination and
  observing geometry into account and calculates the heat conduction
  into the surface. The non-isotropic emission, caused by the distribution
  of surface slopes, shadows, multiple scattering and mutual heating,
  is described by a 2-parameter beaming model with $f$, the fraction of
  the surface covered by craters, and $\rho$, the r.m.s.\ of the surface
  slopes. 
  The TPM, as well as the STM, were mainly used for main-belt asteroids.
  It allows detailed thermophysical studies of individual asteroids
  (e.g., M\"uller \& Lagerros \cite{mueller98}) and predicts for well-known
  asteroids the thermal flux with high accuracy (M\"uller \& Lagerros
  \cite{mueller02}).
  M\"uller (\cite{mueller02a}) investigated the capabilities and
  limitations of the TPM in the context of NEA thermal observations.

  Here we present N-band photometric and spectroscopic observations
  (Sect.\ \ref{sec:obs}). The interpretation of our \object{2002~NY40}
  observations was for the first time done with a combined NEATM and TPM 
  approach (Sect.\ \ref{sec:mod}). We discuss the results of our analysis 
  in Sect.~\ref{sec:discussion}. Using the case of \object{2002~NY40} as
  an example, we also demonstrate the possibilities and limitations of
  applying thermal models to asteroids during close Earth fly-bys.
  
%_______________________________________________________________________
  
%%%%%%%%%%%%%%%%%%%%%%%%%%%%%%%%%%%%%%%%
%  Observations and Data Reduction     %
%%%%%%%%%%%%%%%%%%%%%%%%%%%%%%%%%%%%%%%%
\section{Observations and Data Reduction}
\label{sec:obs} 

  The \object{2002~NY40} observations were taken with the TIMMI2
  instrument (K\"aufl et al. \cite{kaeufl03}) at the ESO La Silla 3.6\,m
  telescope.

  \begin{table}[h!tb]
    \caption{Summary of TIMMI2 observations of asteroid \object{2002~NY40}
             on August 17th/18th, 2002. The observations on August 17th
	     were taken
	     between airmass 1.2 and 1.3, on August 18th between 1.9 and 2.0.
             r, $\Delta$ and $\alpha$ are given as seen from La Silla.
	     The spectra from August 17th were taken with a
             1.2$^{\prime \prime}$ slit, on August 18th with a
             3$^{\prime \prime}$ slit. \label{tbl:obslog}}
    \begin{tabular}{rllllll}
      \hline
      \hline
      \noalign{\smallskip}
           & \multicolumn{2}{c}{Mid-Time} & Filter & r    & $\Delta$ & $\alpha$     \\
      No   & \multicolumn{2}{c}{(Day UT)} & Band   & [AU] & [AU]     & [$^{\circ}$] \\
      \noalign{\smallskip}
      \hline
      \noalign{\smallskip}
      1 & 17 & 04:48 & Ngrism & 1.025406 & 0.013973 & 21.98  \\        
      2 & 17 & 04:57 & Ngrism & 1.025330 & 0.013900 & 22.06  \\        
      3 & 17 & 05:07 & Ngrism & 1.025245 & 0.013820 & 22.16  \\        
      \noalign{\smallskip}
      4 & 18 & 02:02 & N1     & 1.014650 & 0.004535 & 58.37  \\        
      5 & 18 & 02:08 & Ngrism & 1.014600 & 0.004504 & 58.88  \\        
      6 & 18 & 02:13 & Ngrism & 1.014557 & 0.004478 & 59.30  \\        
      7 & 18 & 02:18 & Ngrism & 1.014515 & 0.004452 & 59.74  \\        
      8 & 18 & 02:23 & Ngrism & 1.014473 & 0.004427 & 60.17  \\        
    \noalign{\smallskip}
    \hline
    \end{tabular}
  \end{table}
  
  For all observations we utilized a standard chopping and nodding
  technique to reduce the atmospheric and telescope background
  emission. Chop and nod throws were 10$^{\prime \prime}$,
  respectively. For the imaging observations, a pixel scale of
  0.2$^{\prime \prime}$ was chosen, and typical on source integration
  times of about 1\,minute for each filter were used.
  Exposure times for the low-resolution spectrum (resolution
  $\sim$160) were between 90 and 220\,sec.

  The weather conditions were never ideal and parts of the two nights
  were affected by clouds. Table~\ref{tbl:obslog} contains only the
  good quality measurements together with the observing geometries.
  
  The photometric N1-band flux was calibrated against measurements of
  HD~187642 and HD~12929.
  Colour differences between stars and \object{2002~NY40} were
  negligible (about 1-3\,\%). For observation No.\ 4 in Table.~\ref{tbl:obslog}
  we obtained through standard aperture photometry a calibrated
  8.6\,$\mu$m flux density of 22.2$\pm$1.6\,Jy.
  
  TIMMI2 spectroscopic data were reduced and analysed using
  procedures described by Siebenmorgen et al.\ (\cite{siebenmorgen04}),
  including an airmass dependent extinction
  correction according to Sch\"utz \& Sterzik (\cite{schuetz04}).
  The absolute flux level of TIMMI2 spectroscopic measurements is
  generally better than 5\,\% (K\"ampgen \& Siebenmorgen \cite{kaempgen04}).
  However, our data are not as reliable mainly due to the fast speed of
  \object{2002~NY40} in combination with small slit sizes.
  Therefore, we normalised the spectra from August 18th to the N1
  flux density. The spectra from the first day were
  normalised to a model flux (see Sect.~\ref{sec:mod}).  The spectra are
  shown in Figs.~\ref{fig:spec1mod} and \ref{fig:spec2mod}. The
  spectral shape is reliable in the range between about 8 and 13\,$\mu$m.

%_______________________________________________________________________

%%%%%%%%%%%%%%%%%%%%%%%%%%%%
%  Thermal Modelling       %
%%%%%%%%%%%%%%%%%%%%%%%%%%%%
\section{Thermal Modelling}
\label{sec:mod}

  \begin{figure}[h!]
    \begin{center}
      \rotatebox{90}{\resizebox{!}{\hsize}{\includegraphics{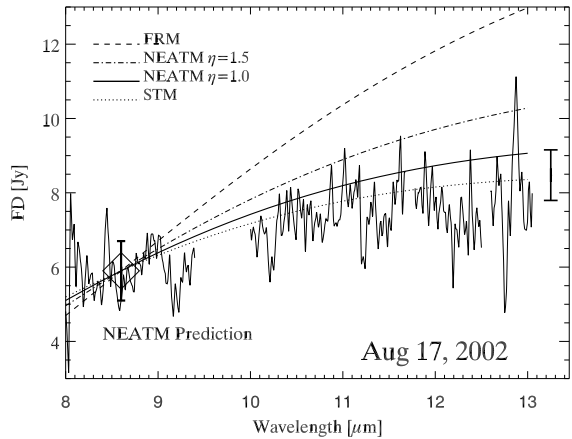}}}
      \caption{The best N-band spectrum of \object{2002~NY40} from August 17th, 2002
               (No.\ 3 in Table~\ref{tbl:obslog}). It was spectroscopically
               calibrated against HD\,133774 and against HD\,187642 and
	       then averaged. The 1\,$\sigma$-error bar represents the
	       signal processing error together with the 1\,$\sigma$-average
	       error (on the right side).
               The residual atmospheric features (Ozone around 9.58$\pm$0.3\,$\mu$m
               and CO$_2$ around 11.73 and 12.55\,$\mu$m) were cut out.
               The spectra were normalised to a NEATM ($\eta=1.0$)
               prediction at 8.6\,$\mu$m for the given observing epoch.
       \label{fig:spec1mod}}
    \end{center}
  \end{figure}

  \begin{figure}[h!]
    \begin{center}
      \rotatebox{90}{\resizebox{!}{\hsize}{\includegraphics{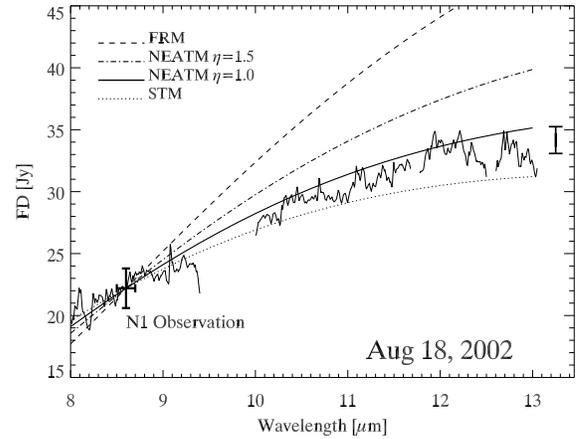}}}
      \caption{The average of the two best N-band spectra
                of \object{2002~NY40} on August 18th, 2002 (No.\ 5 and 6
                in Table~\ref{tbl:obslog}), both calibrated against
                HD\,187642. The error bar was calculated
                in the same way as for Fig.~\ref{fig:spec1mod}.
                The residual atmospheric features were cut out and 
                the spectra were normalised to the N1-band photometric flux.
       \label{fig:spec2mod}}
    \end{center}
  \end{figure}

  We applied the NEATM to determine radiometric diameter and albedo values
  from our N1-band photometric flux. For the beaming value we used
  $\eta=1.5$, as recommended by Delb\'o et al.\ ({\cite{delbo03}) for
  phase angles larger than 45$^{\circ}$. As asteroid intrinsic 
  input parameters we took an absolute magnitude of $H=19.6\pm0.2$\,mag and
  a slope parameter $G=0.15\pm0.20$,
  based on an estimated extrapolation from 21 lightcurve observations
  between end of July and August 16th, 2002. The H-estimates for August
  17th are much more accurate, but unfortunately our thermal photometry
  was not reliable for that day. Our observations of August 18th took
  place about 1\,hour after the lightcurve minimum (J.\ Licandro,
  personal communication).
  
  The constraints from the optical and thermal measurements can
  be graphically shown (e.g., Sekiguchi et al.\ \cite{sekiguchi03})
  with respect to the NEATM assumptions and formulae (e.g., Delb\'o
  \& Harris \cite{delbo02}).
  Our NEATM analysis resulted in an effective diameter of 0.27$\pm$0.01\,km
  and an albedo value of $0.34+0.06/-0.05$ (Fig.~\ref{fig:neatm_DpV}).

  \begin{figure}[h!]
    \begin{center}
      \rotatebox{90}{\resizebox{!}{\hsize}{\includegraphics{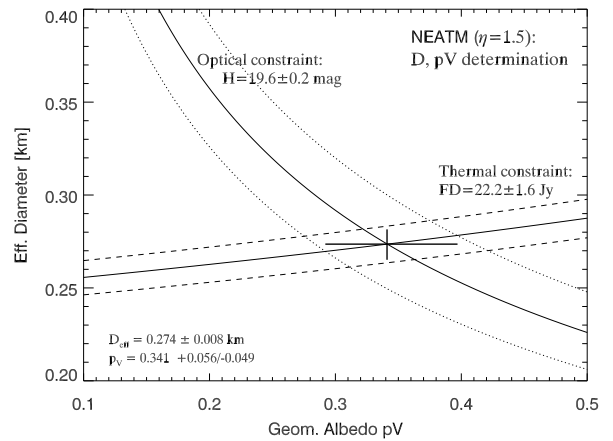}}}
      \caption{NEATM analysis of the N1-band photometric flux. The H-estimate
               is relatively uncertain and propagates mainly into the
	       uncertainty of $p_V$, while the $D_{eff}$ uncertainty is dominated
	       by the thermal flux which is relatively accurate. The $D_{eff}$
	       and $p_V$ results are connected to the assumption that the
	       model does not introduce a systematic error.
       \label{fig:neatm_DpV}}
    \end{center}
  \end{figure}

  Based on these values, we determined a NEATM 8.6\,$\mu$m flux density of
  5.9 $\pm$ 0.8\,Jy for the August 17th spectra, which we used for normalisation
  (see Fig.~\ref{fig:spec1mod}). This NEATM prediction is based
  on the Delb\'o et al. (\cite{delbo03}) recommendation to take
  a value of $\eta=1.0$ for data taken at phase angles below 45$^{\circ}$.

  For both days we calculated
  spectral energy distributions for the mean epochs of the N-band
  spectra through the STM, the Fast Rotating Model (FRM; e.g., Lebofsky \&
  Spencer \cite{lebofsky89}) and the NEATM with $\eta=1.0$ and
  $\eta=1.5$.  A nice summary of all formulae can be found in
  Delb\'o \& Harris (\cite{delbo02}).  The model predictions are
  shown in Figs.~\ref{fig:spec1mod} and \ref{fig:spec2mod}.

  We also applied the TPM by Lagerros
  (\cite{lagerros96}; \cite{lagerros97}; \cite{lagerros98})
  to our observations.
  We assumed a spherical asteroid with an equator-on viewing
  angle under the given phase angle and a prograde rotation
  with $P_{sid}=19.98$\,hours (Pravec et al.\ \cite{pravec04}).
  M\"uller et al.\ (\cite{mueller99}) derived $f=0.6$
  and $\rho=0.7$ for large regolith-covered main-belt asteroids.
  In addition, wee included a relatively smooth surface with a low
  crater coverage ($f=0.4$, $\rho=0.4$) and a very rough surface
  with 100\,\% crater coverage ($f=1.0$, $\rho=1.0$), similar
  to the studies by Dotto et al. (\cite{dotto00}) and
  M\"uller (\cite{mueller02a}). For the emissivity, we used a
  constant, wavelength-independent value of 0.9.

  The thermal behaviour of asteroids is strongly coupled to the
  thermal inertia $\Gamma$, in combination with the orientation
  of the body in space, its rotation period and the sense of
  rotation. A high $\Gamma$ combined with a rapid rotation
  (sub-solar and sub-observer points on the equator)
  would lead to a surface temperature
  distribution described by the FRM. A non-rotating, low $\Gamma$
  object would correspond to the STM temperature distribution.
  Here, we varied $\Gamma$ between 10 (regolith covered surface
  at about 100\,K) and 2500\,$\mathrm{J\,m^{-2}\,s^{-0.5}\,K^{-1}}$
  (solid rock; Jakosky \cite{jakosky86}).

  \begin{table}[h!tb]
    \caption{The radiometric diameter and albedo values for the
             different models based on N-band photometry.
             The values are discussed in Sect.~\ref{sec:discussion}.
             \label{tbl:modres}}
    \begin{tabular}{llll}
      \hline
      \hline
      \noalign{\smallskip}
      Model  &  Diameter [km] & Albedo & Remarks \\
      \noalign{\smallskip}
      \hline
      \noalign{\smallskip}
      STM         &      0.20$\pm$0.02  &      0.64$^{+0.09}_{-0.08}$  & $\eta=0.756$ \\
      FRM         &      0.30$\pm$0.02  &      0.29$^{+0.05}_{-0.04}$  & \\
      {\bf NEATM} & {\bf 0.27$\pm$0.01} & {\bf 0.34$^{+0.06}_{-0.05}$} & {\bf $\eta=1.5$} \\
      NEATM       &      0.22$\pm$0.01  &      0.52$^{+0.08}_{-0.07}$  & $\eta=1.0$ \\
      {\bf TPM}   & {\bf 0.28$\pm$0.03} & {\bf 0.35$\pm$0.02}      & {\bf $\Gamma=1000$} \\
      TPM         &      0.23$\pm$0.03  &      0.51$\pm$0.03  & $\Gamma=100$ \\
    \noalign{\smallskip}
    \hline
    \end{tabular}
  \end{table}

  In order to obtain a similar diameter and albedo value as produced
  by the NEATM calculations, we had to use a high $\Gamma$ value of
  about 1000\,$\mathrm{J\,m^{-2}\,s^{-0.5}\,K^{-1}}$.
  With these values we obtained a TPM diameter of 0.28$\pm$0.03\,km
  and an albedo of 0.35$\pm$0.02. The errors include the uncertainties in
  the measured flux density, the given H-G variations, the above
  specified range of ($\rho$,f)-values and a 20\,\% variation in $\Gamma$.
  
  In Table~\ref{tbl:modres}, we summarise the radiometric diameter and
  albedo values for different models applied to observation No.\ 4
  (Table~\ref{tbl:obslog}). As a combined NEATM/TPM result we took
  $D_{eff}=0.28\pm0.03$ and $p_V=0.34\pm0.06$.

  \begin{figure}[h!]
    \begin{center}
      \rotatebox{90}{\resizebox{!}{\hsize}{\includegraphics{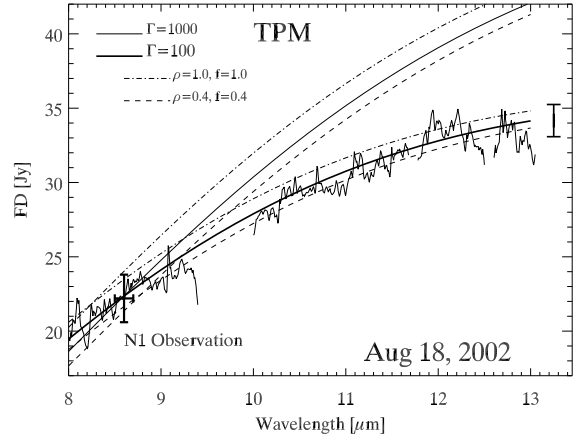}}}
      \caption{N-band spectrum of \object{2002~NY40} from August 18th, 2002,
               with the error bar as in Fig.~\ref{fig:spec2mod}.
               TPM predictions with $\Gamma=1000$ (upper solid line)
               and $\Gamma=100$\,$\mathrm{J\,m^{-2}\,s^{-0.5}\,K^{-1}}$
               (lower solid line) are shown. The dashed and dashed-dotted lines
               indicate the influence of surface roughness expressed
               through $\rho$ and $f$ for both values of $\Gamma$,
	       respectively. Default values are $f=0.6$ and $\rho=0.7$.
       \label{fig:spec2_tpm}}
    \end{center}
  \end{figure}

  Figure~\ref{fig:spec2_tpm} shows the TPM solution for the
  August 18th spectra with
  $\Gamma=1000$\,$\mathrm{J\,m^{-2}\,s^{-0.5}\,K^{-1}}$ and
  $\Gamma=100$\,$\mathrm{J\,m^{-2}\,s^{-0.5}\,K^{-1}}$.
  For both models the influences of a relatively smooth
  and a very rough surface are shown as dashed and dashed-dotted lines.
  The corresponding figure for the August 17th spectrum is very similar.
  The $\Gamma=100$ solution  matches nicely, the $\Gamma=1000$ solution
  is steeper and at long wavelengths significantly above the observed
  spectrum.

%_______________________________________________________________________
  
%%%%%%%%%%%%%%%%%%%%%%%%%%%%
%        Discussion        %
%%%%%%%%%%%%%%%%%%%%%%%%%%%%
\section{Discussion}	\label{sec:discussion}
%_______________________________________________________________________

  Our analysis was guided by the radar size estimate of about
  250$\times$420\,m (E.\ Howell, personal communication) and the radiometric
  results (Rivkin et al.\ \cite{rivkin03}) of a 290-420\,m diameter
  and an albedo preference of about 0.25.
  All model calculations (see Table~\ref{tbl:modres}) produced diameters
  below or at the lower end of these numbers, while the albedo
  was in all cases 0.29 or larger.

  The STM of Lebofsky et al.\ (\cite{lebofsky86}) has a phase angle
  correction of $\beta_E=0.01$\,mag\,deg$^{-1}$, which is considered
  to be a good approximation for phase angles $\le$\,30$^{\circ}$
  (Morrison \cite{morrison77}). For our photometry at 59$^{\circ}$
  it is therefore not a good option. The resulting diameter and
  albedo values are extreme and far away from the radar size or
  the albedo of a Q-class asteroid. The STM fit to the spectral slope 
  is astoundingly good, only for the August 18th data the STM prediction
  falls slightly below the NEATM prediction.

  The FRM led to an acceptable agreement with the radar data.
  But the FRM did not match the spectral shape of the two spectroscopic
  data sets (Figs.\ \ref{fig:spec1mod} and \ref{fig:spec2mod}).

  The NEATM ($\eta=1.5$) and the TPM ($\Gamma=1000$) produced solutions
  which are comparable to the radar size, but again they cannot match
  the observed SEDs.
  On the other hand, a NEATM with $\eta=1.0$ or a TPM with
  $\Gamma=100$ match the spectra, but produce an unrealistically small
  diameter of about 220-230\,m and a very high albedo above 0.50.
  We do not have a convincing explanation for this, but with only
  one reliable photometric flux it is difficult to find a solution,
  unless \object{2002~NY40} turns out to be a small, very high
  albedo object. It might also be that strong deviations from
  a spherical shape seen under the large phase angle of
  58.37$^{\circ}$ cause thermal effects which are difficult
  to model. On the other hand, one set of model parameters fits
  all spectra, although they were taken at very different phase
  angles. This is true for the NEATM and the TPM.

  The NEATM 8.6\,$\mu$m prediction for the spectra of the first day are
  5.9\,Jy ($\eta=1.0$) and 3.6\,Jy ($\eta=1.5$). The corresponding TPM
  calculations (taking also the  NEATM diameter and albedo) give
  6.8\,Jy ($\Gamma=100$) and 4.1\,Jy ($\Gamma=1000$).
  A reliable photometry for this day would therefore indicate which
  of the models has the better simulation of the thermal phase angle
  effects.

  We cannot confirm a phase angle trend on the beaming parameter
  as proposed by Delb\'o et al.\ (\cite{delbo03}). A NEATM solution
  with $\eta=1.0$ fits the spectra of both days at phase angles
  of about 22$^{\circ}$ and 59$^{\circ}$.
  
  The $\Gamma$-value is very crucial for our TPM analysis. A value of
  $\Gamma=15$\,$\mathrm{J\,m^{-2}\,s^{-0.5}\,K^{-1}}$, as it was used
  for regolith-covered main-belt asteroids (M\"uller et al.\ \cite{mueller99}),
  leads to a very small diameter and an albedo of about 0.6. Taking the
  lunar soil value of $\Gamma=39$\,$\mathrm{J\,m^{-2}\,s^{-0.5}\,K^{-1}}$
  (Keihm \cite{keihm84}) does not improve the situation much.
  Only large $\Gamma$-values in the order of
  1000\,$\mathrm{J\,m^{-2}\,s^{-0.5}\,K^{-1}}$
  give reasonable results. However, such a large value can only be the case if
  \object{2002~NY40} has a surface with no or little regolith.
  A dust layer on the surface would always produce much
  lower thermal inertias. 

  Through a TPM analysis it is also possible to investigate the sense of
  rotation (e.g., M\"uller \cite{mueller02a}). We assumed a prograde
  rotator meaning that the terminator is coming from the cold night side
  of the asteroid on August 18th. Taking a retrograde rotator, with the terminator
  coming from the warm day side, would require a $\Gamma$-value of
  about 3000\,$\mathrm{J\,m^{-2}\,s^{-0.5}\,K^{-1}}$ to obtain a diameter
  similar to the radar diameter. Such value would be greater than the
  value for solid rock (Jakosky \cite{jakosky86})
  and therefore not very likely. Again,
  with only one reliable flux it is difficult to obtain a final
  conclusion, but the TPM analysis points to a prograde rotator
  with a cold terminator after opposition. This result will also be useful
  in case of a possible ambiguous solution of its pole orientation
  in the future.

  The TPM curves in Fig.~\ref{fig:spec2_tpm} demonstrate that the
  beaming model, expressed in $\rho$ and $f$, has no big
  influence on the 8 to 13\,$\mu$m predictions. Distinguishing
  between a very rough, cratered surface with pronounced thermal
  beaming effect and a relatively smooth surface with no enhanced
  radiation at small phase angles was therefore not possible for our
  data set. But the beaming does influence the flux differences
  between two phase angles (see also M\"uller \cite{mueller02a}).
  Thermal observations with a good coverage
  in phase angle in combination with a few additional spectra would
  give the possibility to determine $\Gamma$ and indicate if the 
  asteroid has a rough, cratered and therefore old surface or a
  relatively smooth, young surface.
  
  Rivkin et al.\ (\cite{rivkin03}) identified olivine and pyroxene in
  the near-IR spectra. The most diagnostic feature in the N-band
  is the Christiansen peak. It is associated with the principal
  molecular vibration band where the refractive index changes 
  rapidly. For silicates it appears generally as peak between
  7.5 and 9.5\,$\mu$m (Dotto et al.\ \cite{dotto02}).
  We did not see any prominent increase in emittance in this range.
  The feature might have been just outside the reliable wavelength
  range or the quality of the spectra was not sufficient to identify
  silicate features.

%_______________________________________________________________________
  
%%%%%%%%%%%%%%%%%%%%%%%%%%%%
%        Conclusion        % 
%%%%%%%%%%%%%%%%%%%%%%%%%%%%
\section{Conclusion}    \label{sec:conclusion}
%_______________________________________________________________________

  Guided by the radar size, we found that both the NEATM with $\eta=1.5$
  and the TPM with $\Gamma=1000\pm200$\,$\mathrm{J\,m^{-2}\,s^{-0.5}\,K^{-1}}$
  produce consistent effective diameter values of 0.28$\pm$0.03\,km for the
  time of the N-band photometric observation.
  The corresponding geometric albedo is 0.34$\pm$0.06.
  
  Neither of the two model solutions give a satisfying match to the 
  spectra obtained in two observing sessions, separated by slightly more
  than one rotation period. A match of the spectra required either a
  NEATM with $\eta=1.0$ or a TPM with a thermal inertia in the order
  of 100\,\,$\mathrm{J\,m^{-2}\,s^{-0.5}\,K^{-1}}$. No phase angle
  dependence of the NEATM $\eta$-parameter can be seen in our data.

  The TPM analysis excluded a dust-covered surface and pointed towards
  a bare rock surface or a very coarse regolith.
  We obtained reasonable values for the thermal inertia, the
  diameter and albedo, but only assuming a prograde rotation.
  This means that we saw a cold terminator during the observations
  which took place when \object{2002~NY40} was just after opposition,
  but before its closest Earth-approach.

  The example of \object{2002~NY40} shows the different possibilities
  to analyse thermal observations of NEAs through the NEATM and the
  TPM. A carefully planned observing campaign under photometric
  conditions and covering a wide range in solar phase angle
  (without large changes in the aspect angles) are ideal for this
  kind of investigations. High quality photometric fluxes are the key
  to distinguish between different model approaches, but a
  good coverage in visual lightcurves is also required to determine
  H-/G-values and the orientation of the object at the time of
  the thermal observations. Additional calibrated N-band spectra
  allow for estimates of the thermal inertia to prove the existence of a
  dust regolith and the sense of rotation.
  The August 17th/18th 2002 fly-by of \object{2002~NY40} was in principle
  an ideal case for such a study. Only better atmospheric conditions for
  mid-IR observations and a dedicated campaign in the optical would
  have provided more accurate results.
  Close fly-bys at distances below 0.01\,AU are quite frequent
  (the MPC lists currently 14 such event for the period January to
  May 2004), but most of these events are only known on short notice
  or after the fly-by. In fact, there is not even one such
  event listed the period June 2004 to December 2009.
  The next close Earth approach of \object{2002~NY40}
  will be on July 26th, 2005, but at a relatively large
  distance of 0.43\,AU.
    
\begin{acknowledgements}
%   We thank 2002~NY40 for missing the Earth.
   We would like to thank A.\ W.\ Harris (DLR) and M.\ Delb\'o
   for their support in reproducing their NEATM model code.
\end{acknowledgements}
%_______________________________________________________________________

%%%%%%%%%%%%%%%%%%%%%%%%%
%       Bibliography    %
%%%%%%%%%%%%%%%%%%%%%%%%%

%_______________________________________________________________________


\begin{thebibliography}{}
\bibitem[1989]{bowell89}
         Bowell, E., Hapke, B., Domingue, D.\ et al.\ 1989,
         in Asteroids II,
         R.\ P.\ Binzel, T.\ Gehrels \& M.\ S.\ Matthews (Eds.)
	 Arizona University Press, 524
\bibitem[2002]{clark02}
         Clark, B.\ E., Hapke, B., Pieters, C.\ \& Britt, D.\ 2002,
         in Asteroids III, W.\ F.\ Bottke Jr., A.\ Cellino, P.\ Paolicchi,
         \& R.\ P.\ Binzel (Eds), University of Arizona Press, Tucson, p.585-599
\bibitem[1977]{cruikshank77}
         Cruikshank, D.\ P.\ \& Jones, T.\ J.\ 1977,
         Icarus 31, 427
\bibitem[2002]{delbo02}
         Delb\'o, M.\ \& Harris, A.\ W.\ 2002,
         Meteoritics \& Planetary Science 37, 1929
\bibitem[2003]{delbo03}
         Delb\'o, M., Harris, A.\ W., Binzel, R.\ P., Pravec, P.\
         \& Davies, J.\ K.\ 2003,
         Icarus 166, 116
\bibitem[2000]{dotto00}
         Dotto, E., M\"uller, T.\ G., Barucci, M.\ A.\ et al.\ 2000,
         A\&A 358, 1133
\bibitem[2002]{dotto02}
         Dotto, E., Barucci, M.\ A., M\"uller, T.\ G.\ et al.\ 2002,
	 A\&A 393, 1065
\bibitem[1998]{harris98}
         Harris A.\ W.\ 1998.
         Icarus 131, 291
\bibitem[1998]{harris98a}
         Harris, A.\ W., Davies, J.\ K.\ \& Green, S.\ F.\ 1998,
         Icarus 135, 441
\bibitem[1999]{harris99}
         Harris A.\ W.\ \& Davies J.\ K.\ 1999.
         Icarus 142, 464
\bibitem[2003]{howell03}
         Howell, E.\ S., Rivkin, A.\ S., Nolan, M.\ C.\ et al.\ 2003,
         Abstract book of the IAU Conference, Sydney, July 2003,
         JD 19, Abstract No.\ 1872, 254
\bibitem[1986]{jakosky86}
         Jakosky, B.\ M.\ 1986,
         Icarus 66, 117
\bibitem[2004]{kaempgen04}
         K\"amgen, K.\ \& Siebenmorgen, R.\ 2004,
         in High Resolution Infrared Spectroscopy in Astronomy,
         H.\ U.\ K\"aufl, R.\ Siebenmorgen \& A.\ Moorwood (Eds.),
         Springer-Verlag, in press
\bibitem[2003]{kaeufl03}
         K\"aufl, H.\ U., Sterzik, M.\ \& Siebenmorgen, R.\ 2003,
         SPIE 4841, 117
\bibitem[1984]{keihm84}
         Keihm, S.\ J.\ 1984,
         Icarus 60, 568
\bibitem[1996]{lagerros96}
         Lagerros, J.\ S.\ V.\ 1996, A\&A 310, 1011
\bibitem[1997]{lagerros97}
         Lagerros, J.\ S.\ V.\ 1997, A\&A 325, 1226
\bibitem[1998]{lagerros98}
         Lagerros, J.\ S.\ V.\ 1998, A\&A 332, 1123
\bibitem[1979]{lebofsky79}
         Lebofsky, L.\ A., Lebofsky, M.\ J.\ \& Rieke, G.\ H.\ 1979,
	 AJ 84, 885
\bibitem[1986]{lebofsky86}
         Lebofsky, L.\ A., Sykes, M.\ V., Tedesco, E.\ F.\ et al.\ 1986,
	 Icarus 68, 239
\bibitem[1989]{lebofsky89}
         Lebofsky, L.\ A.\ \& Spencer, J.\ R.\ 1989,
         in Asteroids II,
         R.\ P.\ Binzel, T.\ Gehrels \& M.\ S.\ Matthews (Eds.)
	 Arizona University Press, 128
\bibitem[1976]{morrison76}
         Morrison, D.\ 1976,
         Icarus 28, 125
\bibitem[1977]{morrison77}
         Morrison, D.\ 1977,
         ApJ 214, 667
\bibitem[1979]{morrison79}
         Morrison, D.\ \& Lebofsky, L.\ 1979, 184
         in Asteroid, T.\ Gehrels (Ed.),
         University of Arizona Press, Tucson,
\bibitem[1998]{mueller98}
         M\"uller, T.\ G.\ \& Lagerros, J.\ S.\ V.\ 1998,
         A\&A 338, 340
\bibitem[1999]{mueller99}
         M\"uller, T.\ G., Lagerros, J.\ S.\ V., Burgdorf, M.\ et al.\ 1999,
	 ESA SP-427, in The Universe as Seen by ISO,
	 P.\ Cox \& M.\ F.\ Kessler (Eds.), 141
\bibitem[2002]{mueller02}
         M\"{u}ller, T.\ G.\ \& Lagerros, J.\ S.\ V.\ 2002,
         A\&A 381, 324
\bibitem[2002]{mueller02a}
         M\"uller, T.\ G.\ 2002,
         M\&PS 37, 1919
\bibitem[2004]{pravec04}
         Pravec, P., Harris, A.\ W., Scheirich, P.\ et al.\ 2004,
         Icarus 2004, submitted
\bibitem[2003]{rivkin03}
         Rivkin, A.\ S., Howell, E.\ S., Bus, S\. J.\ et al.\ 2003,
	 34th Annual Lunar and Planetary Science Conference March 2003,
	 Abstract No.\ 1722, http://www.mit.edu/~asrivkin/lpsc03B.pdf
\bibitem[2004]{schuetz04}
         Sch\"utz, O.\ \& Sterzik, M.\ 2004,
         in High Resolution Infrared Spectroscopy in Astronomy,
         H.\ U.\ K\"aufl, R.\ Siebenmorgen \& A.\ Moorwood (Eds.),
         Springer-Verlag, in press (astro-ph 0404200)
\bibitem[2003]{sekiguchi03}
         Sekiguchi, T., Abe, M., B\"ohnhard, H.\ et al.\ 2003,
	 A\&A 397, 325
\bibitem[2004]{siebenmorgen04}
         Siebenmorgen, R., Kr\"ugel, E.\ \& Spoon, H.\ W.\ W.\ 2004,
         A\&A 414, 123
\bibitem[2002]{tedesco02}
         Tedesco, E.\ F., Noah, P.\ V., Noah, M.\ \& Price, S.\ D.\ 2002,
         AJ 123, 1056
\bibitem[1989]{veeder89}
         Veeder, G.\ J., Hanner, M.\ S., Matson, D.\ L.\ et al.\ 1989,
	 AJ 97, 1211
\end{thebibliography}
\end{document}